\documentclass[aip,jcp,reprint,twocolumn,superscriptaddress,showpacs,floatfix]{revtex4-1}

\usepackage[latin9]{inputenc}
\usepackage{bm}
\usepackage{amsmath}
\usepackage{graphicx}
\usepackage{epsfig}
\usepackage{mathrsfs}
\usepackage{graphicx}
\usepackage{amssymb}
\usepackage{multirow}
\usepackage{color}
\usepackage{dcolumn}
\usepackage{bm}
\usepackage{float}
\usepackage{placeins}
\definecolor{red}{rgb}{1,0,0}

\makeatletter
\@ifundefined{textcolor}{}
{%
 \definecolor{BLACK}{gray}{0}
 \definecolor{WHITE}{gray}{1}
 \definecolor{RED}{rgb}{1,0,0}
 \definecolor{GREEN}{rgb}{0,1,0}
 \definecolor{BLUE}{rgb}{0,0,1}
 \definecolor{CYAN}{cmyk}{1,0,0,0}
 \definecolor{MAGENTA}{cmyk}{0,1,0,0}
 \definecolor{YELLOW}{cmyk}{0,0,1,0}
}

\makeatother

\begin{document}
\title{Designing Circle Swimmers: Principles and Strategies }
\author{Zhiyu Cao}
\author{Huijun Jiang}
\author{Zhonghuai Hou}
\thanks{E-mail: hzhlj@ustc.edu.cn}
\affiliation{Department of Chemical Physics \& Hefei National Laboratory for Physical
Sciences at Microscales, iChEM, University of Science and Technology
of China, Hefei, Anhui 230026, China}
\date{\today}
\begin{abstract}
Various microswimmers move along circles rather than straight lines
due to their swimming mechanisms, body shapes or hydrodynamic effects.
Here, we adopt the concepts of stochastic thermodynamics to analyze
circle swimmers confined in a two-dimensional plane, and study the
trade-off relations between various physical quantities such as precision,
energy cost and rotational speed. Based on these findings, we predict
principles and strategies for designing microswimmers of special optimized
functions under limited energy resource conditions, which will bring
new experimental inspiration for designing smart motors.
\end{abstract}
\pacs{05.40.-a, 05.70.Ln, 02.50.Ey}
\maketitle

\section{Introduction}

Microswimmers are intrinsically far away from equilibrium by continuously
transferring internal or external energy into their mechanical motion
\citep{menzel2015tuned,solon2015active,weber2011active,debnath2016diffusion,ten2015can}.
From biological motors, bacteria to synthetic active colloids, a large
amount of microswimmers have received great attentions due to their
nontrivial dynamical behaviors forbidden in equilibrium systems \citep{ramaswamy2010mechanics,romanczuk2012active,vicsek2012collective,marchetti2013hydrodynamics,elgeti2015physics,bechinger2016active}.
Due to their self-propulsion motions, microswimmers serve as good
candidates for cargo-carriers in the realm of natural or man-made
micro-machines, which are of significant interest for applications
such as drug delivery, biosensing, or shuttle-transport of living
cells and emulsion droplets \citep{raz2008efficiency,gutman2016optimizing,ma2015catalytic,demirors2018active,debnath2018activated},
to list a few.

In particular, circle swimmers moving in curved trajectories have
been widely applied in cargo delivery\citep{van2009clockwise,reichhardt2013dynamics,kummel2013circular,yang2014self,ao2015diffusion,schirmacher2015anomalous,geiseler2016chemotaxis,jahanshahi2017brownian,lowen2019active,su2019tunable,su2019disordered}.
For instance, the sperm-flagella driven Micro-Bio-Robot (MBR) has been reported to perform precisely point-to-point closed-loop motion under
the influence of external magnetic field with applications towards
targeted drug delivery and microactuation \citep{khalil2014biocompatible,magdanz2013development,magdanz2016dynamic}.
Synthetic catalytic bimetallic nanomotors have been engineered to
pick up, haul, and release micrometer-scale cargo with constant velocity
circular movements to mimic nanoscale biomotors in biological
systems \citep{fournier2005synthetic,marine2013diffusive,takagi2013dispersion}.
Magnetic swimmers actuated by externally applied field, rotating at
a constant speed in a plane, have been envisaged for targeted drug
delivery \citep{buzhardt2021magnetically}, and so on. Since such
microswimmers generally function at small scales, their transport
dynamics are intrinsically stochastic. Therefore, how to work against
such inherent fluctuations with good performance is of great importance
for their designs \citep{jahanshahi2017brownian,liebchen2019optimal,van2008dynamics,muratore2014nanomechanical,nemoto2019optimizing,pietzonka2019autonomous}.

\begin{figure}
\begin{centering}
\includegraphics[width=1\columnwidth]{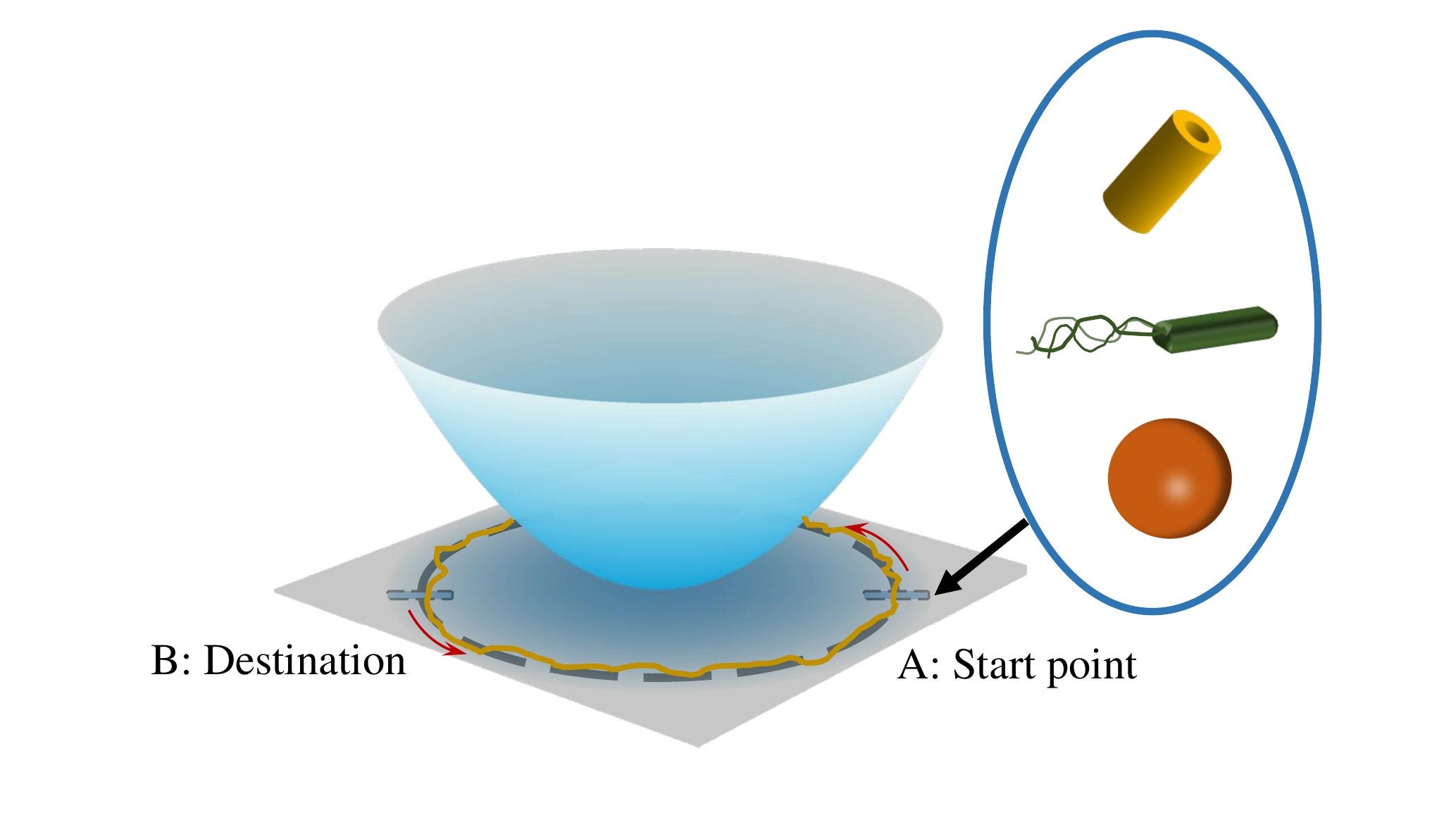}
\par\end{centering}
\caption{Schematic sketch of cargo delivering by various types of Brownian
circle swimmers. The circle swimmers deliver the cargo from A (start
point) to B (destination) along a fluctuating circular orbit (the
black line). One may control the swimmer motion by changing its active
torque for circular motion, active force for translational motion
or by applying possible external potentials (blue surface). The yellow
curved line represents the actual stochastic trajectory, and the black
dashed line represents the averaged trajectory.}

\label{fig:1}
\end{figure}

Here, we address such an issue by studying a general model of Brownian
circle swimmer in a 2D plane (see Fig.\ref{fig:1}), subjected to
an active force with constant amplitude $F_{0}$, an internal torque
$M$ and an external harmonic potential with strength $k$. By using
the framework of stochastic thermodynamics \citep{seifert2012stochastic,seifert2008stochastic,seifert2019stochastic,jarzynski2011equalities},
we have analyzed the heat dissipation of the system as a function
of $F_{0}$, $M$ and $k$, identified as thermodynamic cost of the
swimmer. Interestingly, we find that there exists an optimal torque
such that the cyclic cost $Q_{cyc}$ could reach a minimum value,
which serves as a lower bound for the cost to sustain a stable circular
motion. In addition, a trade-off relation between the precision of
the circular motion and the thermodynamic cost has been found, indicating
that an increased energy supply is needed to maintain punctual transport
process as expected. By utilizing the thermodynamic uncertainty relations(TURs)
\citep{barato2015thermodynamic}, we have also obtained an analytical
expression for the transport efficiency $\chi_{\theta}$, which properly
quantifies the performance of the swimmer to work with large speed,
high precision, but low cost. With these results, we are able to propose
proper strategies for designing swimmers under certain constraints,
for instance, when the cyclic energy supply $Q_{0}$ is limited. Analysis
shows that the maximum $\chi_{\theta}$ increases with increasing
$Q_{0}$, which can be further enhanced by increasing the external
potential strength $k$. Finally, we illustrate our predicted design
principles and strategies by numerical simulations.

\section{Model}

As shown in Fig.\ref{fig:1}, we consider a microswimmer performing
circular motion, carrying cargo along a fluctuating circular orbit
in two-dimensional $xy$ plane from $A$ (the start point) to $B$
(the destination) and then going back, which completes a single cyclic
transport task. These microswimmers can be natural or synthetic, such
as bacteria, rods, or spherical Janus particles, $etc$ \citep{van2008dynamics},
and related scenarios have been reported in environmental remediation,
targeted drug delivery or transportation of cells \citep{ma2015catalytic}.

For simplicity, we model the circle swimmer by an active spherical
particle of radius $R$. The particle is subjected to a self-propelled
force with amplitude $F_{0}$ along the orientation $\bm{e}=(\cos\varphi,\sin\varphi)$
and an internal torque $\Omega$ (along the $z$ direction) that drives
the circular motion. Without loss of generality, we also consider
that the particle is controlled by an external conservative force
$\bm{F}_{k}\left(\mathbf{r}\right)=-\partial_{\bm{\mathbf{r}}}U_{k}\left(\bm{\mathbf{r}}\right)$
with $U_{k}\left(\bm{\mathbf{r}}\right)$ the external potential ($k$
is a control parameter) and $\bm{r}=(x,y)$ the center-of-mass position
of the particle. Neglecting hydrodynamic effects, the particle dynamics
can be described by the following overdamped Langevin equations:

\begin{equation}
\gamma_{t}\dot{\bm{r}}=\bm{F}_{k}+F_{0}\bm{e}+\bm{\xi},\label{eq:1}
\end{equation}
\begin{equation}
\dot{\varphi}=\Omega+\zeta,\label{eq:2}
\end{equation}
where $\gamma_{t}$ is the translational friction coefficient. For
simplicity, we have set the rotational friction coefficient $\gamma_{r}=1$
throughout the paper, thus the angular velocity of swimmer equals
to the value of torque $\Omega$. The fluctuation terms, $\bm{\xi}=\left\{ \xi_{x},\xi_{y}\right\} $
and $\zeta$, are both independent Gaussian white noises with zero
mean, satisfying $\left\langle \bm{\xi}(t)\bm{\xi}(s)\right\rangle =2\mathbf{I}\gamma_{t}T\delta(t-s)$
and $\left\langle \zeta(t)\zeta(s)\right\rangle =2T\delta(t-s)$,
where $T$ is the ambient temperature and $\mathbf{I}$ the unit tensor.
According to \textcolor{red}(the fluctuation-dissipation relation), the short-time translational
and rotational diffusion coefficients of the particle read $D_{t}=T/\gamma_{t}$
and $D_{r}=T$ (where we have set Boltzmann constant $k_{B}=1$),
with $D_{t}/D_{r}=4R^{2}/3$ holding for a spherical particle \citep{jahanshahi2017brownian}.
The corresponding Fokker-Planck equation for the probability density
function $P(\bm{r},\varphi;t)$ is given by
\begin{equation}
\partial_{t}P(\bm{r},\varphi;t)=-\partial_{\bm{r}}\cdot\bm{J}{}_{\bm{r}}(\bm{r},\varphi;t)-\partial_{\varphi}J_{\varphi}(\bm{r},\varphi;t),\label{eq:FPE}
\end{equation}
where the probability currents read
\[
\bm{J}_{\bm{r}}(\bm{r},\varphi;t)=\gamma_{t}^{-1}\left(\bm{F}+F_{0}\bm{e}-T\partial_{\bm{r}}\right)P(\bm{r},\varphi;t)
\]
and
\[
J_{\varphi}(\bm{r},\varphi;t)=\left(\Omega-T\partial_{\varphi}\right)P(\bm{r},\varphi;t)
\]
respectively. Hereafter, we mainly focus upon the steady state with
$\partial_{t}P_{ss}\left(\mathbf{r},\varphi\right)=0$.

To deliver cargo with good performance for the swimmer considered
here, it may be required that the swimmer can reach the destination
with a high speed and precisely on time, but with relatively low dissipation
\citep{hwang2018energetic}. Therefore, fast delivery with high accuracy
and low cost can be viewed as a design principle of the swimmer.

\section{Thermodynamic cost}

To establish a quantitative measure of such a principle, firstly,
one needs to investigate the cost of the delivery process. Since the
swimmer follows stochastic dynamics, here we use the strategy of stochastic
thermodynamics \citep{seifert2012stochastic}, which has been widely
studied in the last decade. Note that the stochastic thermodynamic has been successfully generalized to active matter systems \citep{fodor2016far,kyriakopoulos2016leading,pearce2016linear,mandal2017entropy,pietzonka2017entropy,shankar2018hidden,dabelow2019irreversibility,pietzonka2019autonomous,bonilla2019active,nemoto2019optimizing,burkholder2019fluctuation,szamel2019stochastic}. For instance, Pearce and Giomi have studied the linear response to leadership, effective temperature and decision making for a collection of flocking agents \citep{pearce2016linear}. Kyriakopoulos $et.al$ have investigated the nonequilibrium response of polar ordered active fluid to external alignments \citep{kyriakopoulos2016leading}. Besides, Burkholder and Brady have obtained a generazlied fluctuation-dissipation relationship for active suspensions with time-reversal symmetry broken \cite{burkholder2019fluctuation}. More recently, Szamal have proposed a generalization of the standard stochastic thermodynamics to self-propelled particles \cite{szamel2019stochastic}.

The framework of stochastic thermodynamics allows us to calculate the average entropy production
and then heat dissipation rate $\Sigma$ along a delivery cycle, which
can be identified as the thermodynamic cost for the delivery process. Following Seifert's spirit \citep{seifert2012stochastic}, one can
define a trajectory-based entropy for the the microswimmer as $s\left(t\right)=-\ln P\left(\bm{r},\varphi;t\right)$,
the change rate of which can be written as
\begin{equation}
\dot{s}\left(t\right)=\dot{s}_{tot}\left(t\right)-\dot{s}_{m}\left(t\right)\label{eq:s}
\end{equation}
where
\begin{equation}
\dot{s}_{m}\left(t\right)=\left[\left(\bm{F}+F_{0}\bm{e}\right)\cdot\dot{\bm{r}}+\Omega\dot{\varphi}\right]/T\label{eq:sm}
\end{equation}
is the so called medium entropy production \citep{szamel2019stochastic}
that can be related directly to the heat dissipation rate $\dot{q}$
via $\dot{q}=T\dot{s}_{m}\left(t\right)$. The term $\dot{s}_{tot}\left(t\right)=\dot{s}\left(t\right)+\dot{s}_{m}\left(t\right)=-\partial_{t}\ln P\left(\bm{r},\varphi;t\right)+\text{\ensuremath{T^{-1}\gamma_{t}\bm{v}_{r}\left(\bm{r},\varphi;t\right)\cdot\dot{\bm{r}}}}+\text{\ensuremath{T^{-1}v_{\varphi}\left(\bm{r},\varphi;t\right)\dot{\varphi}}}$
is the change rate of the trajectory-dependent total entropy production,
where $\bm{v}_{r}=\bm{J}_{r}/P$ and $v_{\varphi}=J_{\varphi}/P$
are the local mean velocities \citep{seifert2012stochastic}. Upon
averaging over trajectories, one can obtain

\begin{align}
\dot{S}_{tot} & =\left\langle \dot{s}_{tot}\right\rangle \nonumber \\
 & =T^{-1}\int d\bm{r}d\varphi\left(\gamma_{t}\bm{v}_{r}^{2}+v_{\varphi}^{2}\right)P\left(\bm{r},\varphi;t\right)\ge0.\label{eq:Stot}
\end{align}

In the steady state, we can use $\Sigma=T\left\langle \dot{s}_{m}\right\rangle =T\dot{S}_{tot}$
to calculate the heat dissipation rate, i.e., the rate of thermodynamic
cost for the delivery process. To be specific, we now set the external
potential $U\left(\bm{r}\right)=k\bm{r}^{2}/2$, where the potential
strength $k$ works as a control parameter. For further purpose, it
is convenient to rewrite Eq.\ref{eq:1} in polar coordinates ($\mathbf{r}=re^{i\theta}$):
\begin{equation}
\dot{r}=-\beta D_{t}kr+\beta D_{t}F_{0}\cos\left(\theta-\varphi\right)+\sqrt{2D_{t}}\xi_{r},\label{eq:dr-1}
\end{equation}

\begin{equation}
\dot{\theta}=-\frac{\beta D_{t}F_{0}}{r}\sin\left(\theta-\varphi\right)+\sqrt{2D_{t}}\xi_{\theta},\label{eq:dth-1}
\end{equation}
where $\xi_{r}=\cos\theta\xi_{x}+\sin\theta\xi_{y}$, $\xi_{\theta}=-\left(\text{\ensuremath{\sin\theta}\ensuremath{\xi_{x}-\cos\theta}\ensuremath{\xi_{y}}}\right)/r$.
In the steady state, the angles $\theta$ and $\varphi$ would be
phase-locked, i.e., $\theta-\varphi\simeq\text{const}$ and $\left\langle \dot{\theta}\right\rangle \simeq\Omega$.
The circle swimmer will rotate along a stochastic limit cycle, and
the mean radius of the circular orbit can be computed by setting $\beta D_{t}kr_{m}\simeq\beta D_{t}F_{0}\cos\left(\theta-\varphi\right)$
\citep{jahanshahi2017brownian}, i.e.,
\begin{equation}
r_{m}=\frac{F_{0}}{\sqrt{\gamma_{t}^{2}\Omega^{2}+k^{2}}},\label{eq:rm}
\end{equation}
where we have used $-\beta D_{t}F_{0}\sin\left(\theta-\varphi\right)/r_{m}\simeq\Omega$.
Using Eq.\ref{eq:Stot}, one can then obtain the cost rate in the steady
state as (see Appendix A for details)
\begin{align}
\Sigma & \simeq\gamma_{t}\omega^{2}r_{m}^{2}+\omega^{2}\nonumber \\
 & =\frac{F_{0}^{2}\Omega^{2}}{\gamma_{t}\Omega^{2}+\gamma_{t}^{-1}k^{2}}+\Omega^{2}.\label{eq:W}
\end{align}

\section{Optimal torque for cyclic cost}

From Eq.\ref{eq:W}, one can see that the cost rate depends on all
the parameters, the active force $F_{0}$, the torque $\Omega$,
the external force strength $k$, as well as the friction coefficients
$\gamma_{t}$. In general, larger activity and torque will lead to
larger cost rate, while stronger control (larger $k$) will reduce
the cost rate. Nevertheless, due to the circular motion of the swimmer,
a more relevant quantity in real processes would be the cyclic cost
$Q_{cyc}$, i.e., the thermodynamic cost per delivery cycle, which
is given by

\begin{align}
Q_{cyc} & =\left(\frac{2\pi}{\Omega}\right)\Sigma\nonumber \\
 & =\frac{2\pi}{\Omega}\left(\frac{F_{0}^{2}\Omega^{2}}{\gamma_{t}\Omega^{2}+\gamma_{t}^{-1}k^{2}}+\Omega^{2}\right).\label{eq:DeltaW}
\end{align}

It can be seen that, while smaller torque $\Omega$ leads to smaller
cost rate $\Sigma$, it also causes longer cyclic process. An interesting
result is then there exists a trade-off between speed and thermodynamic
cost for the circular motion. In particular, an optimal torque $\Omega_{opt}$
exists for the circle swimmer to minimize the cost per cycle, which
can be obtained by setting $dQ_{cyc}/dM=0$, giving

\begin{equation}
\Omega_{opt}^{2}=\frac{-(\gamma_{t}^{-1}F_{0}^{2}+2\gamma_{t}^{-2}k^{2})+\sqrt{\gamma_{t}^{-2}F_{0}^{4}+8F_{0}^{2}k^{2}\gamma_{t}^{-3}}}{2}.\label{opt-2}
\end{equation}
Note that the condition for the existence of the optimal torque is
that the active and external forces must satisfy $k\le k_{0}=\sqrt{\gamma_{t}F_{0}^{2}}$,
and the minimum $Q_{cyc}^{opt}$ provides a lower bound of the energy
to sustain the circular motion. In the simple case when the external
potential is absent ($k=0$), the optimal torque reads $\Omega_{opt}=F_{0}/\sqrt{\gamma_{t}}$,
and the corresponding minimum cyclic cost reads $Q_{cyc}^{opt}=4\pi F_{0}/\sqrt{\gamma_{t}}$
. With increasing $k$, it can be observed from Eq.\ref{eq:DeltaW}
that the minimum cyclic cost can be further reduced. If lowering $Q_{cyc}$
is the target for the delivery process, Eqs.(\ref{eq:DeltaW}) and
(\ref{opt-2}) serve as the formula for design principles.

\section{Precision-cost trade-off relation}

Besides thermodynamic cost, another important aspect of the cyclic
delivery process is the precision. Similar to biochemical oscillation
processes, the precision can be conveniently measured by the phase
diffusion constant $D_{\theta}=\left(\left\langle \theta^{2}\right\rangle -\left\langle \theta\right\rangle ^{2}\right)/2t$.
Using similar method as in our previous work\citep{cao2020design},
we can obtain
\begin{equation}
D_{\theta}\simeq\frac{D_{t}}{r_{m}^{2}}=\frac{T\left(\gamma_{t}\Omega^{2}+\gamma_{t}^{-1}k^{2}\right)}{F_{0}^{2}}.\label{eq:Dth}
\end{equation}
Combining this result with Eq.(\ref{eq:W}), we reach an interesting
trade-off relation between phase diffusion and dissipation rate, which
reads,
\begin{equation}
D_{\theta}=\frac{T\text{\ensuremath{\Sigma_{c}}}}{\Sigma-\text{\ensuremath{\Sigma_{c}}}},\label{tor1}
\end{equation}
where $\Sigma_{c}=\Omega^{2}$ can be identified as the minimum dissipation
rate of a circular motion.

A few remarks can be made. Firstly, large dissipation rate is required
to reach high precision, which exactly tells the trade-off between
cost and gain. Secondly, for a circle swimmer carrying cargo, the
dissipation rate needs to exceed a critical value $\ensuremath{\Sigma_{c}}$
to transport along the rotational orbit, which is already indicated
in Eq.(\ref{eq:W}), and extra dissipation could be applied to ensure
the swimmer works with a more punctual manner. Thirdly, the inverse
law, Eq.\ref{tor1}, between phase diffusion constant $D_{\theta}$
and dissipation rate $\Sigma$ has also been found in biochemical
oscillation systems \citep{cao2015free,cao2020design}, highlighting
its ubiquitous importance in living systems.

\section{Transport efficiency}

To further characterize the performance and design principle of the
circle swimmer, here we apply the so-called thermodynamic uncertainty
relation (TUR), which was first proposed by Barato and Seifert \citep{barato2015thermodynamic}
in Markovian jump processes, and has been extensively studied recently
\citep{horowitz2019thermodynamic,gingrich2016dissipation,pietzonka2016universal,pigolotti2017generic,proesmans2017discrete,dechant2018entropic,van2019uncertainty,hasegawa2019fluctuation}.
According to Dechant and Sasa \citep{dechant2018current}, a TUR relation
holds for general Langevin systems as $\left(v_{j}^{st}\right)^{2}/\dot{S}_{tot}D_{j}\leqslant1$,
wherein $v_{j}^{st}$ and $D_{j}$ are steady state velocity and diffusivity
of a generalized current variable $j$, respectively. Note that such
a relation implies that one cannot obtain an arbitrary large speed
without either accepting large phase diffusion or investing a large
dissipation, which exactly tells the trade-off among speed, precision
and energy cost. Direct application of the TUR to our present system,
choosing $j$ as $\theta$, is that
\[
\chi_{\theta}=\left(v_{\theta}^{st}\right)^{2}/\dot{S}_{tot}D_{\theta}\le1
\]
where $\chi_{\theta}$ is the transport efficiency \citep{dechant2018current,hwang2018energetic}.
Clearly, $\chi_{\theta}$ can be used to quantitatively describe the
ability for the swimmer to maintain a high-speed, accurate transport
with lower dissipation. Combining Eq.\ref{eq:W} and \ref{eq:Dth},
noting that $v_{\theta}^{st}=\Omega$, we obtain the theoretical expression
of the transport efficiency for the circle swimmer as

\begin{equation}
\chi_{\theta}=\frac{F_{0}^{2}}{F_{0}^{2}+\gamma_{t}\Omega^{2}+\gamma_{t}^{-1}k^{2}}.\label{tur1}
\end{equation}
which is clearly less than 1 for nonzero $\Omega$ or $k$. Obviously,
a larger self-propelled force will increase the transport efficiency,
while a larger torque or potential strength will reduce it.

\begin{figure}
\begin{centering}
\includegraphics[width=0.8\columnwidth]{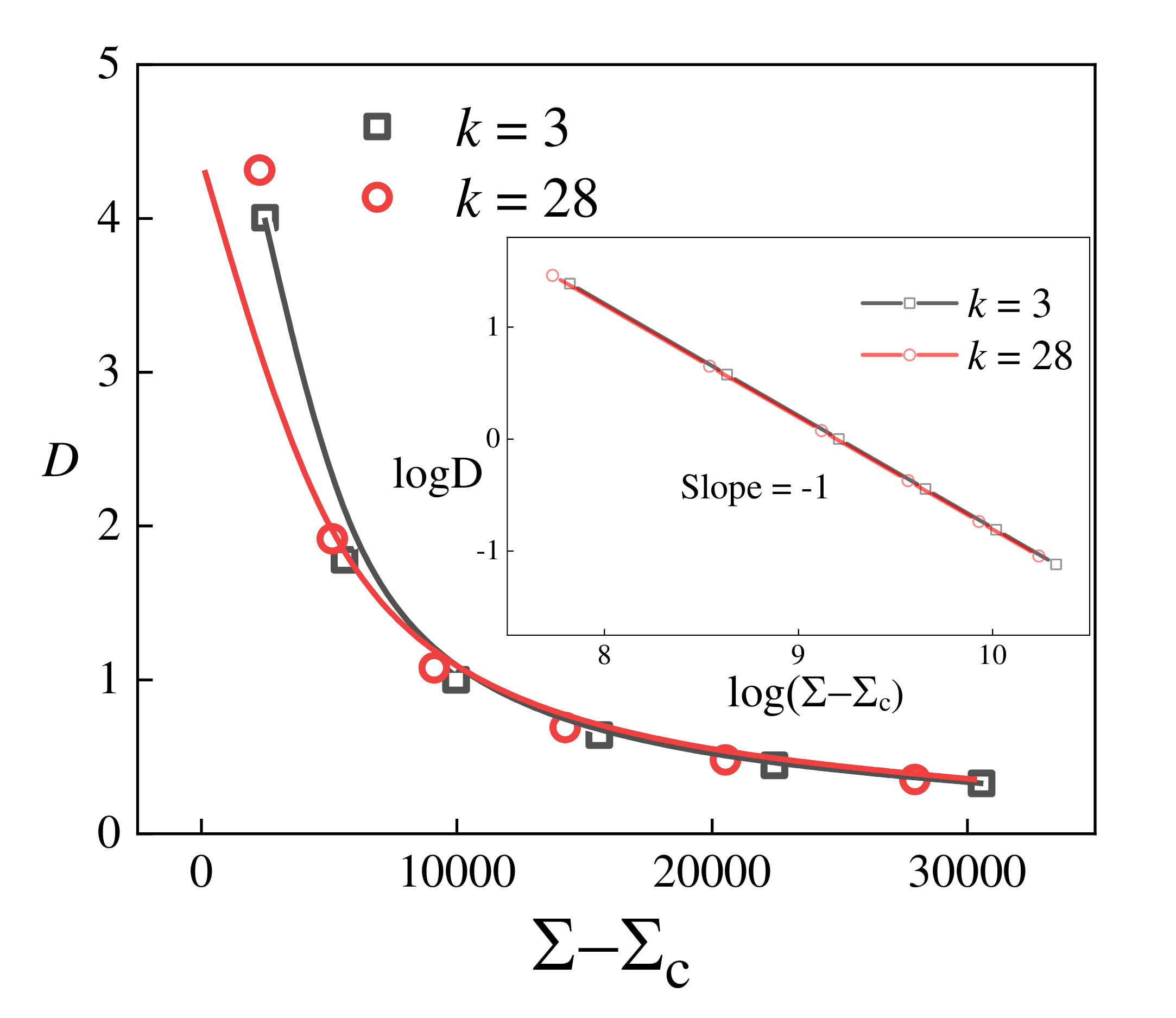}
\par\end{centering}
\caption{(a) The trade-off relation Eq.\ref{tor1} between the normalized phase
diffusion constant $D=D_{\theta}/T$ and the dissipation rate $\Sigma$.
The symbols are obtained from numerical results and solid curves are
the theoretical expressions. The data plotted in $\log$-$\log$ scale
are fitted well by the scaling form $D=\text{\ensuremath{\Sigma_{c}}}\left(\Sigma-\text{\ensuremath{\Sigma_{c}}}\right)^{\alpha}$
with fitting parameter $\alpha=-1$(inset).}

\label{fig:2}
\end{figure}

\section{Design strategy for limited energy resource}

For better performance of the swimmer regarding cargo delivery, it
is suggested that $\chi_{\theta}$ should be as large as possible.
To this goal, it seems that one can simply enhance the activity ($F_{0}$)
and reduce the torque ($\Omega$) or external potential ($k$). However,
as discussed above, increasing $F_{0}$ will also increase the thermodynamic
cost rate $\Sigma$ or the cyclic cost $Q_{cyc}$. In real systems,
a more relevant question would be how to design the circle swimmer
when the energy resource is limited. For instance, when the chemical
substances driving the directional movement are exhausted, the microswimmers
may fail to self-propelled \citep{fournier2005synthetic}. Here, we
focus on a situation when the cyclic supply of energy is limited by
a given value $Q_{0}$, and discuss how to maximize the transport
efficiency $\chi_{\theta}$ under such a constraint. Note that to
sustain the rotation motion, $Q_{0}$ must be larger than the minimum
value $Q_{cyc}^{opt}$ given by Eq.(\ref{eq:DeltaW}). If not, a feasible
way is to increase the external potential strength $k$ and thus reduce
$Q_{cyc}^{opt}$ as indicated also in Eq.(\ref{eq:DeltaW}). A simple
analysis shows that setting the strength $k\ge\sqrt{\gamma_{t}\left(\frac{2\pi F_{0}^{2}\Omega}{Q_{0}-2\pi\Omega}-\gamma_{t}\Omega^{2}\right)}$
can make the swimmer maintain circular motion in the case when $Q_{0}<4\pi F_{0}/\sqrt{\gamma_{t}}$
.

Given that $Q_{0}>4\pi F_{0}/\sqrt{\gamma_{t}}$, circular motion
can be sustained in the absence of external potential, i.e., $k=0$.
For now, the design principle is to maximize the transport efficiency
$\chi_{\theta}=F_{0}^{2}/\left(F_{0}^{2}+\gamma_{t}\Omega^{2}\right)$
under the constraint that $Q_{cyc}\left(F_{0},\Omega\right)=2\pi\left(\frac{F_{0}^{2}}{\gamma_{t}\Omega}+\Omega\right)\le Q_{0}$.
The conditional highest transport efficiency, which is achieved when
$Q_{cyc}=Q_{0}$, can be obtained as (see Appendix B for details)

\begin{equation}
\chi_{\theta,\text{max}}=\left[1+\ensuremath{\frac{4}{\left(C+\sqrt{C^{2}-4}\right)^{2}}}\right]^{-1}\label{eq:chimax-2}
\end{equation}
with $C=\sqrt{\gamma_{t}}Q_{0}/\left(2\pi F_{0}\right)$. Then for
fixed self-propelled force $F_{0}$, one can get higher transport
efficiency with increasing input energy per-cycle $Q_{0}$, and correspondingly
the swimmer works with a torque $\Omega$ determined by the condition
$Q_{cyc}\left(F_{0},\Omega\right)=Q_{0}$. These serve as a reasonable
design strategy for the swimmer, although how to transfer the input
$Q_{0}$ to torque $\Omega$ may require specific technic.

As mentioned above, the lower bound for $Q_{cyc}$ can be decreased
with increasing $k$. Therefore, for given $Q_{0}$, $Q_{0}/Q_{cyc}^{opt}$
will increase with $k$, i.e., the energy supply becomes more abundant
with increasing $k$. Intuitively, the transport efficiency would
become larger if all $Q_{0}$ is transferred to the torque. Unfortunately,
it is not possible to get an analytical expression for $\chi_{\theta,\text{max}}$
for $k\ne0$. By numerical simulation, however, we indeed find that
the maximum transport efficiency can be further improved by increasing
$k$ under the condition $Q_{cyc}\le Q_{0}$, as illustrated below.

\begin{figure}
\begin{centering}
\includegraphics[width=1\columnwidth]{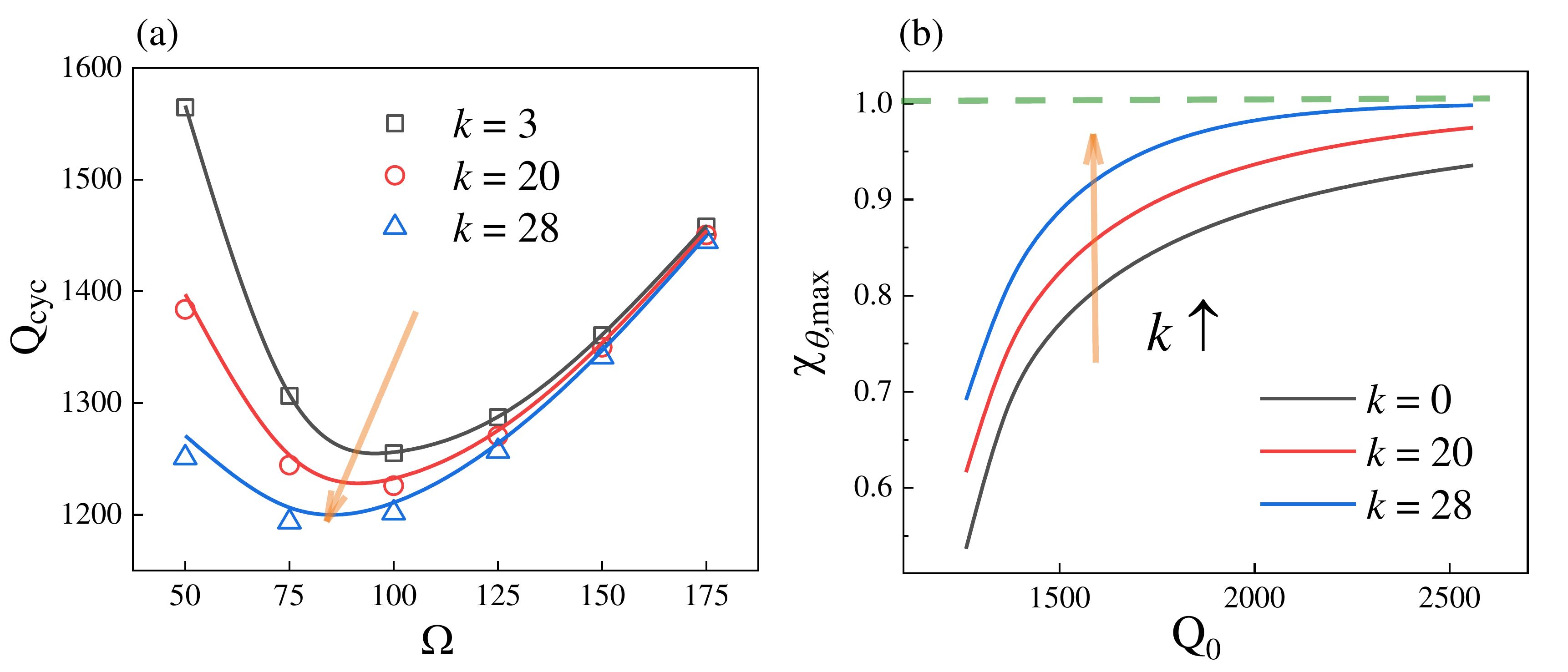}
\par\end{centering}
\caption{(a) Cyclic cost $Q_{cyc}$ as a function of the torque $\Omega$ for
different values of the potential strength $k$. The optimal torque
$\Omega_{opt}$ to minimize $Q_{cyc}$ moves to the left, and the
cyclic energy cost can be further reduced by increasing the potential
strength. Lines: theory. Symbols: simulations. (b) The maximal transport efficiency $\chi_{\theta,\text{max}}$
as a function of the limited energy supply $Q_{0}$ for different
values of the potential strength $k$. The maximal transport efficiency
can also be enhanced by increasing the potential strength. The data is obtained from maximizing the analytical results for transport efficiency $\chi_{\theta}=F_{0}^2/(F_{0}^2+\gamma_{t}\Omega^{2}+\gamma_{t}^{-1}k^2)$ under the constraint $Q_{cyc}\le Q_0$. $F_{0}=100$.}

\label{fig:3}
\end{figure}

\section{Numerical simulations}

To test our previous theoretical predictions, we numerically solve
the Langevin equation \ref{eq:1} and \ref{eq:2} with a time step
$10^{-3}$. The dissipation rate $\Sigma$ is numerically calculated
by using $\Sigma=T\left\langle \dot{s}_{m}\right\rangle $, i.e.,
Eq.(\ref{eq:sm}), and $D_{\theta}$ is computed according to the
definition. The precision-dissipation trade-off relation, Eq.(\ref{tor1}),
has been shown in Fig.\ref{fig:2}. It can be observed that the normalized
phase diffusion constant $D=D_{\theta}/T$ is inversely proportional
to the dissipation rate $\Sigma-\Sigma_{c}$, and the data plotted
in $\log$-$\log$ scale are fitted very well by a line with slope
-1, in good agreement with the theoretical results.

In Fig.\ref{fig:3}(a), we show how the cyclic energy cost $Q_{cyc}$
changes with the internal torque $\Omega$, and the optimal torque
$\Omega_{opt}$ can be clearly identified. As discussed above, the
cyclic energy cost can be further reduced by increasing the potential
strength $k$. At last, for different values of the potential strength
$k$, the maximal transport efficiency $\chi_{\theta,\text{max}}$
under limited energy resource $Q_{0}$ is shown in Fig.\ref{fig:3}(b)
for fixed $F_{0}=100$. Clearly, $\chi_{\theta,\text{max}}$ increases
with $Q_{0}$ and approaching the upper bound 1 for very large energy
input $Q_{0}$, and introducing external potential will further improve
the maximal transport efficiency, in consistent with our previous
predictions.

\section{Conclusion}

In summary, we have studied the stochastic thermodynamics for the
Brownian circle swimmer in a 2D plane. We theoretically obtain an
expression of the cost rate $\Sigma$, and an optimal torque for minimal
cyclic cost $Q_{cyc}$ has been found, which provides the lower bound
for the thermodynamic cost to sustain a stable circular motion. Meanwhile,
it has been revealed that extra energy supply could be applied to
enhance the precision of the transport process by a derived trade-off
relation. Using the general principles for dissipative processes,
the TURs, we also analyze the transport efficiency of cargo transport,
which offers quantitative insight into the performance of the swimmer.
Furthermore, we predict design strategies under the condition when
the cyclic energy resource is limited. Analysis shows that the maximum
$\chi_{\theta}$ gradually approaches the upper bound while increasing
the cyclic supply, which can be further enhanced by introducing an
external potential. By using our approach, it is possible to design
the swimmer for other important conditions, which is left for further
investigation in future work.

Recently, the collective effects in active systems have attracted much attention \citep{farage2015effective}. According to our main result (Eq.\ref{tur1}), the transport efficiency reads as $\chi_{\theta}=(v_{\theta}^{st})^{2}/\dot{S}_{tot}D_{\theta}=F_{0}^2/(F_{0}^2+\gamma_{t}\Omega^{2}+\gamma_{t}^{-1}k^2)$, where $F_0$ is the self-propelled force, $\Omega$ is the torque and $k$ is the potential strength. The steric interactions may induce an effective larger $k$, which will result a decreased transport efficiency. On the other hand, collective interactions could  make the swimmers synchronize, which suppress the phase diffusion (smaller phase diffusion constant $D_\theta $) to enhance the transport efficiency \citep{2020The,lee2018thermodynamic}. Besides, the macroscopic crowding as a result of the inter-particle interactions may also enhance the transport efficiency by suppressing the torque $\Omega$ \citep{sokolov2012models}. In our opinion, the generalization to collective active systems is not straightforward, which deserves further study. Since it has been shown that various
experimental setups are feasible to realize the swimmer model \citep{jahanshahi2017brownian},
we believe that the predicted principles and strategies presented
here are testable, and can offer guidelines for experiments.
\begin{acknowledgments}
This work is supported by MOST(2018YFA0208702), NSFC (32090044, 21973085,
21833007, 21790350, 21521001).
\end{acknowledgments}

\section*{Data Availability}

The data that support the findings of this study are available within
the article 
and its supplementary material. 

\appendix
\onecolumngrid

\section*{Supplementary information}


\section{Cost rate for the circle swimmer}

We now calculate the cost rate in steady state. Following Eq.(\ref{eq:Stot})
of the main text, the energy dissipation rate $\Sigma=\dot{Q}=T\dot{S}_{tot}$
reads
\[
\Sigma=\int d\bm{r}d\varphi\left(\gamma_{t}\bm{v}_{r}^{2}+\gamma_{r}v_{\varphi}^{2}\right)P_{ss}(\bm{r},\varphi)=\Sigma_{r}+\Sigma_{c}.
\]
Here, $\Sigma_{r}=\gamma_{t}\left\langle \bm{v}_{r}^{2}\right\rangle $
and $\Sigma_{c}=\gamma_{r}\left\langle v_{\varphi}^{2}\right\rangle $
denote the cost rate associated with the rotational motion along the
circular orbit and the self chiral motion. Due to the time scale separation
between the radial motion and the angular/chiral motion, $\Sigma_{r}$
can be calculated by approximating $P_{ss}(\bm{r},\varphi)\simeq P_{ss}(\bm{r})P_{ss}(\varphi)$
and $\left\langle r^{2}\right\rangle \simeq r_{m}^{2}$ \citep{cao2015free,cao2020design}
\begin{equation}
\Sigma_{r}\simeq\gamma_{t}\int\int r^{2}\omega^{2}P_{ss}(r,\theta)\cdot rdrd\theta\simeq\gamma_{t}\omega^{2}\left\langle r^{2}\right\rangle \simeq\gamma_{t}\omega^{2}r_{m}^{2}.
\end{equation}
The explicit expression for $\Sigma_{r}$ reads

\begin{equation}
\Sigma_{r}=\frac{F_{0}^{2}\Omega^{2}}{\gamma_{t}\Omega^{2}+\gamma_{t}^{-1}k^{2}}.
\end{equation}
As $\Sigma_{c}\simeq\Omega^{2}$, the cost rate can be derived as
the form of Eq.(\ref{eq:W}) in the main text.

\section{Design strategies for limited energy resource}

By writing $\rho=\sqrt{\gamma_{t}}\Omega/F_{0}$, the design strategies
emerges from maximizing the transport efficiency $\chi_{\theta}=1/\left(1+\ensuremath{\rho}^{2}\right)$
under the condition that $2\le\rho+\text{\ensuremath{\rho^{-1}}}\le\sqrt{\gamma_{t}}Q_{0}/\left(2\pi F_{0}\right)=C,$
where $C$ is the dissipation normalized by self-propelled force.
By solving for the range of the ratio that satisfies the condition,
$\frac{2}{C+\sqrt{C^{2}-4}}\le\rho\le\frac{C+\sqrt{C^{2}-4}}{2}$,
we can obtain the highest transport efficiency
\begin{equation}
\chi_{\theta,\text{max}}=\frac{1}{1+\ensuremath{\rho_{\text{min}}}^{2}}\label{eq:chimax-1}
\end{equation}
with $\ensuremath{\rho_{\text{min}}}=\frac{2}{C+\sqrt{C^{2}-4}}$.
We find that the available supply need to be fully utilized to enhance
the transport efficiency.

Further, we start to analyze the effect of the external potential.
Similarly, the question emerges from how to maximize the transport
efficiency $\frac{1}{1+F_{0}^{-2}(\Omega^{2}+k^{2})}$ under the condition
$\frac{F_{0}^{2}\Omega}{\Omega^{2}+k^{2}}+\Omega\le\frac{Q_{0}}{2\pi}$.
By simply rewriting, the transport efficiency reads
\begin{equation}
\chi_{\theta}=\left(1+\frac{\Omega}{\frac{Q_{0}}{2\pi}-\Omega}\right)^{-1},
\end{equation}
i.e., the smaller the torque allowed to be selected, the transport
efficiency will be higher. For simplicity, we assume the potential
strength $k$ is relatively small. Approximating by Taylor expansion,
the condition reads $\frac{F_{0}^{2}}{\Omega}+\Omega\lesssim\frac{Q_{0}}{2\pi}+\frac{F_{0}^{2}k^{2}}{\Omega^{3}}$,
which implies that the allowed torque is smaller by increasing the
potential strength, i.e., the maximal transport efficiency can be
further enhanced.

\section{Numerical details}

\begin{figure}
\begin{centering}
\includegraphics[width=0.4\columnwidth]{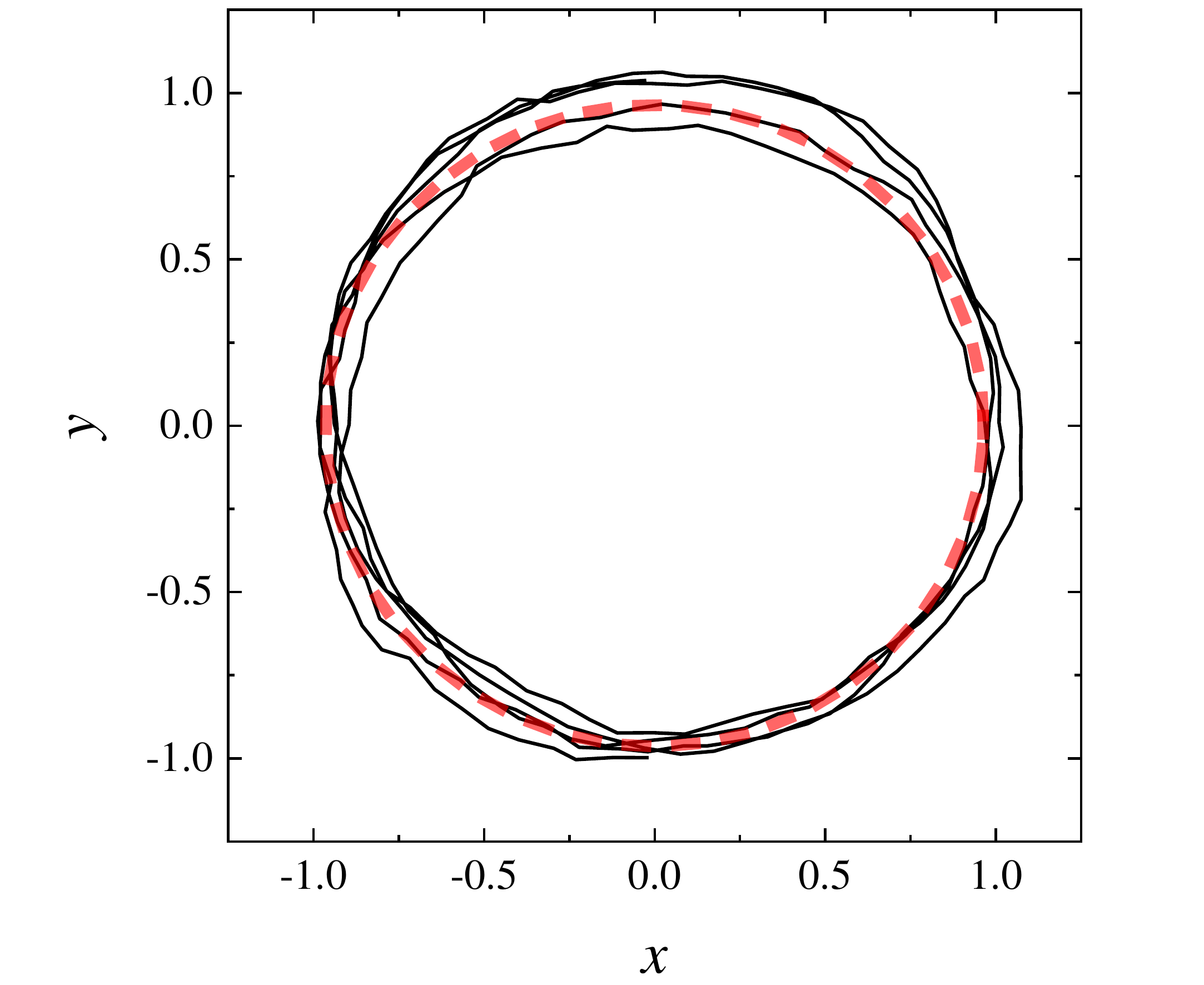}
\par\end{centering}
\renewcommand\thefigure{S1}
\caption{Typical simulated trajectory in the xy plane (black lines) from Eqs.\ref{eq:1} and \ref{eq:2}. Here, $F_0=100$, $\Omega=100$, $k=28$, $T=0.1$ and  $\gamma_t=1$. The red dashed circle represents the stationary circular orbit with the radius $r_{m}=F_{0}/\sqrt{\gamma_{t}^{2}\Omega^{2}+k^{2}}$.}

\label{fig:S1}
\end{figure}

The starting point for the numerical simulations are Eqs.\ref{eq:1} and \ref{eq:2}, which are discretized according to the Euler's method as
\begin{equation}
\gamma_{t}[\bm{r}(t+\Delta t)-\bm{r}(t)]=\bm{F}_{k}(\bm{r}(t)))\Delta t+F_{0}\bm{e}(t)\Delta t+\sqrt{2\gamma_{t}T\Delta t}\mathbf{I}\bm{N}_{\xi},\label{eq:C1}
\end{equation}
\begin{equation}
\varphi(t+\Delta t)-\varphi(t)=\Omega\Delta t+\sqrt{2T\Delta t}N_{\zeta}\label{eq:C2}
\end{equation}
with $\bm{r}=(x,y)$ and $\bm{e}=(\cos\varphi,\sin\varphi)$. $\bm{N}_{\xi}$ and $N_{\zeta}$ are independent and normally distributed random variables with zero mean and unit variance. The discretized time step $\Delta t=10^{-3}$, the translational friction coefficient $\gamma_t=1$, and the statistical data are obtained from averaging over $10^4$ trajectories. Typical trajectory in the xy plane of the numerical results is plotted in Fig.\ref{fig:S1} (also can be seen Movie S1).

The most important quantities to be calculated are the thermodynamic cost rate $\Sigma$ and transport efficiency $\chi_{\theta}$. Here, we use $\Sigma=T\langle \dot{s}_m\rangle=T\dot{S}_{tot}$ to calculate the thermodynamic cost rate. As discussed in the main text, $Tds_m=(\bm{F}+F_0\bm{e})\cdot d\bm{r}+\Omega d\varphi$. Since $d\bm{r}$ and $d\varphi$ can be obtained from the dynamics generating from Eqs.\ref{eq:1} and \ref{eq:2}, $ds_m$ or $\dot{s}_m$ can be calculated numerically. By averaging over trajectories in steady states, the thermodynamic rate $\Sigma$ can then be obtained. Moreover, the transport efficiency $\chi_{\theta}=(v_{\theta}^{st})^2/\dot{S}_{tot}D_{\theta}$, are obtained from the numerical data of $\dot{S}_{tot}$, $v_{\theta}^{st}$ and $D_{\theta}$. As shown above, $\dot{S}_{tot}$ can be obtained from simulations. As $\bm{r}=re^{i\theta}$, $r$ and $\theta$ can be got from the generating dynamics, i.e., $v_{\theta}^{st}=\lim_{t\to \infty}\langle\theta\rangle/t$ and $D_\theta=\lim_{t\to \infty}(\langle\theta^2\rangle-\langle\theta\rangle^2)/2t$ can also be calculated from the steady state trajectories. Therefore, the transport efficiency can be obtained from the simulated data.

\bibliographystyle{amsplain} 

\begin{thebibliography}{{\citenamefont{K{"u}mmel et~al.}(2013)\citenamefont{K{"u}mmel, ten   Hagen, Wittkowski, Buttinoni, Eichhorn, Volpe, L{"o}wen, and   Bechinger}}}
\bibitem[{\citenamefont{Menzel}(2015)}]{menzel2015tuned} \bibinfo{author}{\bibfnamefont{A.~M.}
\bibnamefont{Menzel}}, \bibinfo{journal}{Physics reports} \textbf{\bibinfo{volume}{554}},
\bibinfo{pages}{1} (\bibinfo{year}{2015}).

\bibitem[{\citenamefont{Solon et~al.}(2015)\citenamefont{Solon, Cates, and   Tailleur}}]{solon2015active}
\bibinfo{author}{\bibfnamefont{A.~P.} \bibnamefont{Solon}},
\bibinfo{author}{\bibfnamefont{M.}~\bibnamefont{Cates}},
\bibnamefont{and} \bibinfo{author}{\bibfnamefont{J.}~\bibnamefont{Tailleur}},
\bibinfo{journal}{The European Physical Journal Special Topics}
\textbf{\bibinfo{volume}{224}}, \bibinfo{pages}{1231} (\bibinfo{year}{2015}).

\bibitem[{\citenamefont{Weber et~al.}(2011)\citenamefont{Weber, Radtke,   Schimansky-Geier, and H{"a}nggi}}]{weber2011active}
\bibinfo{author}{\bibfnamefont{C.}~\bibnamefont{Weber}},
\bibinfo{author}{\bibfnamefont{P.~K.} \bibnamefont{Radtke}},
\bibinfo{author}{\bibfnamefont{L.}~\bibnamefont{Schimansky-Geier}},
\bibnamefont{and} \bibinfo{author}{\bibfnamefont{P.}~\bibnamefont{H{"a}nggi}},
\bibinfo{journal}{Physical Review E} \textbf{\bibinfo{volume}{84}},
\bibinfo{pages}{011132} (\bibinfo{year}{2011}).

\bibitem[{\citenamefont{Debnath et~al.}(2016)\citenamefont{Debnath, Ghosh, Li,   Marchesoni, and Li}}]{debnath2016diffusion}
\bibinfo{author}{\bibfnamefont{D.}~\bibnamefont{Debnath}},
\bibinfo{author}{\bibfnamefont{P.~K.} \bibnamefont{Ghosh}},
\bibinfo{author}{\bibfnamefont{Y.}~\bibnamefont{Li}}, \bibinfo{author}{\bibfnamefont{F.}~\bibnamefont{Marchesoni}},
\bibnamefont{and} \bibinfo{author}{\bibfnamefont{B.}~\bibnamefont{Li}},
\bibinfo{journal}{Soft Matter} \textbf{\bibinfo{volume}{12}},
\bibinfo{pages}{2017} (\bibinfo{year}{2016}).

\bibitem[{\citenamefont{Ten~Hagen et~al.}(2015)\citenamefont{Ten~Hagen,   Wittkowski, Takagi, K{"u}mmel, Bechinger, and L{"o}wen}}]{ten2015can}
\bibinfo{author}{\bibfnamefont{B.}~\bibnamefont{Ten~Hagen}},
\bibinfo{author}{\bibfnamefont{R.}~\bibnamefont{Wittkowski}},
\bibinfo{author}{\bibfnamefont{D.}~\bibnamefont{Takagi}},
\bibinfo{author}{\bibfnamefont{F.}~\bibnamefont{K{"u}mmel}},
\bibinfo{author}{\bibfnamefont{C.}~\bibnamefont{Bechinger}},
\bibnamefont{and} \bibinfo{author}{\bibfnamefont{H.}~\bibnamefont{L{"o}wen}},
\bibinfo{journal}{Journal of Physics: Condensed Matter} \textbf{\bibinfo{volume}{27}},
\bibinfo{pages}{194110} (\bibinfo{year}{2015}).

\bibitem[{\citenamefont{Ramaswamy}(2010)}]{ramaswamy2010mechanics}
\bibinfo{author}{\bibfnamefont{S.}~\bibnamefont{Ramaswamy}},
\bibinfo{journal}{Annu. Rev. Condens. Matter Phys.} \textbf{\bibinfo{volume}{1}},
\bibinfo{pages}{323} (\bibinfo{year}{2010}).

\bibitem[{\citenamefont{Romanczuk et~al.}(2012)\citenamefont{Romanczuk,   B{"a}r, Ebeling, Lindner, and Schimansky-Geier}}]{romanczuk2012active}
\bibinfo{author}{\bibfnamefont{P.}~\bibnamefont{Romanczuk}},
\bibinfo{author}{\bibfnamefont{M.}~\bibnamefont{B{"a}r}},
\bibinfo{author}{\bibfnamefont{W.}~\bibnamefont{Ebeling}},
\bibinfo{author}{\bibfnamefont{B.}~\bibnamefont{Lindner}},
\bibnamefont{and} \bibinfo{author}{\bibfnamefont{L.}~\bibnamefont{Schimansky-Geier}},
\bibinfo{journal}{The European Physical Journal Special Topics}
\textbf{\bibinfo{volume}{202}}, \bibinfo{pages}{1} (\bibinfo{year}{2012}).

\bibitem[{\citenamefont{Vicsek and Zafeiris}(2012)}]{vicsek2012collective}
\bibinfo{author}{\bibfnamefont{T.}~\bibnamefont{Vicsek}}
\bibnamefont{and} \bibinfo{author}{\bibfnamefont{A.}~\bibnamefont{Zafeiris}},
\bibinfo{journal}{Physics reports} \textbf{\bibinfo{volume}{517}},
\bibinfo{pages}{71} (\bibinfo{year}{2012}).

\bibitem[{\citenamefont{Marchetti et~al.}(2013)\citenamefont{Marchetti, Joanny,   Ramaswamy, Liverpool, Prost, Rao, and Simha}}]{marchetti2013hydrodynamics}
\bibinfo{author}{\bibfnamefont{M.~C.} \bibnamefont{Marchetti}},
\bibinfo{author}{\bibfnamefont{J.-F.} \bibnamefont{Joanny}},
\bibinfo{author}{\bibfnamefont{S.}~\bibnamefont{Ramaswamy}},
\bibinfo{author}{\bibfnamefont{T.~B.} \bibnamefont{Liverpool}},
\bibinfo{author}{\bibfnamefont{J.}~\bibnamefont{Prost}},
\bibinfo{author}{\bibfnamefont{M.}~\bibnamefont{Rao}}, \bibnamefont{and}
\bibinfo{author}{\bibfnamefont{R.~A.} \bibnamefont{Simha}},
\bibinfo{journal}{Reviews of Modern Physics} \textbf{\bibinfo{volume}{85}},
\bibinfo{pages}{1143} (\bibinfo{year}{2013}).

\bibitem[{\citenamefont{Elgeti et~al.}(2015)\citenamefont{Elgeti, Winkler, and   Gompper}}]{elgeti2015physics}
\bibinfo{author}{\bibfnamefont{J.}~\bibnamefont{Elgeti}},
\bibinfo{author}{\bibfnamefont{R.~G.} \bibnamefont{Winkler}},
\bibnamefont{and} \bibinfo{author}{\bibfnamefont{G.}~\bibnamefont{Gompper}},
\bibinfo{journal}{Reports on progress in physics} \textbf{\bibinfo{volume}{78}},
\bibinfo{pages}{056601} (\bibinfo{year}{2015}).

\bibitem[{\citenamefont{Bechinger et~al.}(2016)\citenamefont{Bechinger,   Di~Leonardo, L{"o}wen, Reichhardt, Volpe, and Volpe}}]{bechinger2016active}
\bibinfo{author}{\bibfnamefont{C.}~\bibnamefont{Bechinger}},
\bibinfo{author}{\bibfnamefont{R.}~\bibnamefont{Di~Leonardo}},
\bibinfo{author}{\bibfnamefont{H.}~\bibnamefont{L{"o}wen}},
\bibinfo{author}{\bibfnamefont{C.}~\bibnamefont{Reichhardt}},
\bibinfo{author}{\bibfnamefont{G.}~\bibnamefont{Volpe}},
\bibnamefont{and} \bibinfo{author}{\bibfnamefont{G.}~\bibnamefont{Volpe}},
\bibinfo{journal}{Reviews of Modern Physics} \textbf{\bibinfo{volume}{88}},
\bibinfo{pages}{045006} (\bibinfo{year}{2016}).

\bibitem[{\citenamefont{Raz and Leshansky}(2008)}]{raz2008efficiency}
\bibinfo{author}{\bibfnamefont{O.}~\bibnamefont{Raz}} \bibnamefont{and}
\bibinfo{author}{\bibfnamefont{A.}~\bibnamefont{Leshansky}},
\bibinfo{journal}{Physical Review E} \textbf{\bibinfo{volume}{77}},
\bibinfo{pages}{055305} (\bibinfo{year}{2008}).

\bibitem[{\citenamefont{Gutman and Or}(2016)}]{gutman2016optimizing}
\bibinfo{author}{\bibfnamefont{E.}~\bibnamefont{Gutman}}
\bibnamefont{and} \bibinfo{author}{\bibfnamefont{Y.}~\bibnamefont{Or}},
\bibinfo{journal}{Physical Review E} \textbf{\bibinfo{volume}{93}},
\bibinfo{pages}{063105} (\bibinfo{year}{2016}).

\bibitem[{\citenamefont{Ma et~al.}(2015)\citenamefont{Ma, Hahn, and   Sanchez}}]{ma2015catalytic}
\bibinfo{author}{\bibfnamefont{X.}~\bibnamefont{Ma}}, \bibinfo{author}{\bibfnamefont{K.}~\bibnamefont{Hahn}},
\bibnamefont{and} \bibinfo{author}{\bibfnamefont{S.}~\bibnamefont{Sanchez}},
\bibinfo{journal}{Journal of the American Chemical Society} \textbf{\bibinfo{volume}{137}},
\bibinfo{pages}{4976} (\bibinfo{year}{2015}).

\bibitem[{\citenamefont{Demir{"o}rs et~al.}(2018)\citenamefont{Demir{"o}rs,   Akan, Poloni, and Studart}}]{demirors2018active}
\bibinfo{author}{\bibfnamefont{A.~F.} \bibnamefont{Demir{"o}rs}},
\bibinfo{author}{\bibfnamefont{M.~T.} \bibnamefont{Akan}},
\bibinfo{author}{\bibfnamefont{E.}~\bibnamefont{Poloni}},
\bibnamefont{and} \bibinfo{author}{\bibfnamefont{A.~R.} \bibnamefont{Studart}},
\bibinfo{journal}{Soft Matter} \textbf{\bibinfo{volume}{14}},
\bibinfo{pages}{4741} (\bibinfo{year}{2018}).

\bibitem[{\citenamefont{Debnath and Ghosh}(2018)}]{debnath2018activated}
\bibinfo{author}{\bibfnamefont{T.}~\bibnamefont{Debnath}}
\bibnamefont{and} \bibinfo{author}{\bibfnamefont{P.~K.} \bibnamefont{Ghosh}},
\bibinfo{journal}{Physical Chemistry Chemical Physics} \textbf{\bibinfo{volume}{20}},
\bibinfo{pages}{25069} (\bibinfo{year}{2018}).

\bibitem[{\citenamefont{van Teeffelen et~al.}(2009)\citenamefont{van Teeffelen,   Zimmermann, and L{"o}wen}}]{van2009clockwise}
\bibinfo{author}{\bibfnamefont{S.}~\bibnamefont{van Teeffelen}},
\bibinfo{author}{\bibfnamefont{U.}~\bibnamefont{Zimmermann}},
\bibnamefont{and} \bibinfo{author}{\bibfnamefont{H.}~\bibnamefont{L{"o}wen}},
\bibinfo{journal}{Soft Matter} \textbf{\bibinfo{volume}{5}},
\bibinfo{pages}{4510} (\bibinfo{year}{2009}).

\bibitem[{\citenamefont{Reichhardt and   Reichhardt}(2013)}]{reichhardt2013dynamics}
\bibinfo{author}{\bibfnamefont{C.}~\bibnamefont{Reichhardt}}
\bibnamefont{and} \bibinfo{author}{\bibfnamefont{C.~O.} \bibnamefont{Reichhardt}},
\bibinfo{journal}{Physical Review E} \textbf{\bibinfo{volume}{88}},
\bibinfo{pages}{042306} (\bibinfo{year}{2013}).

\bibitem[{\citenamefont{K{"u}mmel et~al.}(2013)\citenamefont{K{"u}mmel, ten   Hagen, Wittkowski, Buttinoni, Eichhorn, Volpe, L{"o}wen, and   Bechinger}}]{kummel2013circular}
\bibinfo{author}{\bibfnamefont{F.}~\bibnamefont{K{"u}mmel}},
\bibinfo{author}{\bibfnamefont{B.}~\bibnamefont{ten Hagen}},
\bibinfo{author}{\bibfnamefont{R.}~\bibnamefont{Wittkowski}},
\bibinfo{author}{\bibfnamefont{I.}~\bibnamefont{Buttinoni}},
\bibinfo{author}{\bibfnamefont{R.}~\bibnamefont{Eichhorn}},
\bibinfo{author}{\bibfnamefont{G.}~\bibnamefont{Volpe}},
\bibinfo{author}{\bibfnamefont{H.}~\bibnamefont{L{"o}wen}},
\bibnamefont{and} \bibinfo{author}{\bibfnamefont{C.}~\bibnamefont{Bechinger}},
\bibinfo{journal}{Physical review letters} \textbf{\bibinfo{volume}{110}},
\bibinfo{pages}{198302} (\bibinfo{year}{2013}).

\bibitem[{\citenamefont{Yang et~al.}(2014)\citenamefont{Yang, Qiu, and   Gompper}}]{yang2014self}
\bibinfo{author}{\bibfnamefont{Y.}~\bibnamefont{Yang}},
\bibinfo{author}{\bibfnamefont{F.}~\bibnamefont{Qiu}}, \bibnamefont{and}
\bibinfo{author}{\bibfnamefont{G.}~\bibnamefont{Gompper}},
\bibinfo{journal}{Physical Review E} \textbf{\bibinfo{volume}{89}},
\bibinfo{pages}{012720} (\bibinfo{year}{2014}).

\bibitem[{\citenamefont{Ao et~al.}(2015)\citenamefont{Ao, Ghosh, Li, Schmid,   H{"a}nggi, and Marchesoni}}]{ao2015diffusion}
\bibinfo{author}{\bibfnamefont{X.}~\bibnamefont{Ao}}, \bibinfo{author}{\bibfnamefont{P.~K.}
\bibnamefont{Ghosh}}, \bibinfo{author}{\bibfnamefont{Y.}~\bibnamefont{Li}},
\bibinfo{author}{\bibfnamefont{G.}~\bibnamefont{Schmid}},
\bibinfo{author}{\bibfnamefont{P.}~\bibnamefont{H{"a}nggi}},
\bibnamefont{and} \bibinfo{author}{\bibfnamefont{F.}~\bibnamefont{Marchesoni}},
\bibinfo{journal}{EPL (Europhysics Letters)} \textbf{\bibinfo{volume}{109}},
\bibinfo{pages}{10003} (\bibinfo{year}{2015}).

\bibitem[{\citenamefont{Schirmacher et~al.}(2015)\citenamefont{Schirmacher,   Fuchs, H{"o}fling, and Franosch}}]{schirmacher2015anomalous}
\bibinfo{author}{\bibfnamefont{W.}~\bibnamefont{Schirmacher}},
\bibinfo{author}{\bibfnamefont{B.}~\bibnamefont{Fuchs}},
\bibinfo{author}{\bibfnamefont{F.}~\bibnamefont{H{"o}fling}},
\bibnamefont{and} \bibinfo{author}{\bibfnamefont{T.}~\bibnamefont{Franosch}},
\bibinfo{journal}{Physical review letters} \textbf{\bibinfo{volume}{115}},
\bibinfo{pages}{240602} (\bibinfo{year}{2015}).

\bibitem[{\citenamefont{Geiseler et~al.}(2016)\citenamefont{Geiseler,   H{"a}nggi, Marchesoni, Mulhern, and Savel'ev}}]{geiseler2016chemotaxis}
\bibinfo{author}{\bibfnamefont{A.}~\bibnamefont{Geiseler}},
\bibinfo{author}{\bibfnamefont{P.}~\bibnamefont{H{"a}nggi}},
\bibinfo{author}{\bibfnamefont{F.}~\bibnamefont{Marchesoni}},
\bibinfo{author}{\bibfnamefont{C.}~\bibnamefont{Mulhern}},
\bibnamefont{and} \bibinfo{author}{\bibfnamefont{S.}~\bibnamefont{Savel'ev}},
\bibinfo{journal}{Physical Review E} \textbf{\bibinfo{volume}{94}},
\bibinfo{pages}{012613} (\bibinfo{year}{2016}).

\bibitem[{\citenamefont{Jahanshahi et~al.}(2017)\citenamefont{Jahanshahi,   L{"o}wen, and Ten~Hagen}}]{jahanshahi2017brownian}
\bibinfo{author}{\bibfnamefont{S.}~\bibnamefont{Jahanshahi}},
\bibinfo{author}{\bibfnamefont{H.}~\bibnamefont{L{"o}wen}},
\bibnamefont{and} \bibinfo{author}{\bibfnamefont{B.}~\bibnamefont{Ten~Hagen}},
\bibinfo{journal}{Physical Review E} \textbf{\bibinfo{volume}{95}},
\bibinfo{pages}{022606} (\bibinfo{year}{2017}).

\bibitem[{\citenamefont{L{"o}wen}(2019)}]{lowen2019active} \bibinfo{author}{\bibfnamefont{H.}~\bibnamefont{L{"o}wen}},
\bibinfo{journal}{Physical Review E} \textbf{\bibinfo{volume}{99}},
\bibinfo{pages}{062608} (\bibinfo{year}{2019}).

\bibitem[{\citenamefont{Su et~al.}(2019{\natexlab{a}})\citenamefont{Su, Jiang,   and Hou}}]{su2019tunable}
\bibinfo{author}{\bibfnamefont{J.}~\bibnamefont{Su}}, \bibinfo{author}{\bibfnamefont{H.-J.}
\bibnamefont{Jiang}}, \bibnamefont{and} \bibinfo{author}{\bibfnamefont{Z.-H.}
\bibnamefont{Hou}}, \bibinfo{journal}{The Journal of Physical
Chemistry C} \textbf{\bibinfo{volume}{123}}, \bibinfo{pages}{17624}
(\bibinfo{year}{2019}{\natexlab{a}}).

\bibitem[{\citenamefont{Su et~al.}(2019{\natexlab{b}})\citenamefont{Su, Jiang,   and Hou}}]{su2019disordered}
\bibinfo{author}{\bibfnamefont{J.}~\bibnamefont{Su}}, \bibinfo{author}{\bibfnamefont{H.}~\bibnamefont{Jiang}},
\bibnamefont{and} \bibinfo{author}{\bibfnamefont{Z.}~\bibnamefont{Hou}},
\bibinfo{journal}{Soft Matter} \textbf{\bibinfo{volume}{15}},
\bibinfo{pages}{6830} (\bibinfo{year}{2019}{\natexlab{b}}).

\bibitem[{\citenamefont{Khalil et~al.}(2014)\citenamefont{Khalil, Magdanz,   Sanchez, Schmidt, and Misra}}]{khalil2014biocompatible}
\bibinfo{author}{\bibfnamefont{I.~S.} \bibnamefont{Khalil}},
\bibinfo{author}{\bibfnamefont{V.}~\bibnamefont{Magdanz}},
\bibinfo{author}{\bibfnamefont{S.}~\bibnamefont{Sanchez}},
\bibinfo{author}{\bibfnamefont{O.~G.} \bibnamefont{Schmidt}},
\bibnamefont{and} \bibinfo{author}{\bibfnamefont{S.}~\bibnamefont{Misra}},
\bibinfo{journal}{Journal of Micro-Bio Robotics} \textbf{\bibinfo{volume}{9}},
\bibinfo{pages}{79} (\bibinfo{year}{2014}).

\bibitem[{\citenamefont{Magdanz et~al.}(2013)\citenamefont{Magdanz, Sanchez,   and Schmidt}}]{magdanz2013development}
\bibinfo{author}{\bibfnamefont{V.}~\bibnamefont{Magdanz}},
\bibinfo{author}{\bibfnamefont{S.}~\bibnamefont{Sanchez}},
\bibnamefont{and} \bibinfo{author}{\bibfnamefont{O.~G.} \bibnamefont{Schmidt}},
\bibinfo{journal}{Advanced Materials} \textbf{\bibinfo{volume}{25}},
\bibinfo{pages}{6581} (\bibinfo{year}{2013}).

\bibitem[{\citenamefont{Magdanz et~al.}(2016)\citenamefont{Magdanz, Guix,   Hebenstreit, and Schmidt}}]{magdanz2016dynamic}
\bibinfo{author}{\bibfnamefont{V.}~\bibnamefont{Magdanz}},
\bibinfo{author}{\bibfnamefont{M.}~\bibnamefont{Guix}},
\bibinfo{author}{\bibfnamefont{F.}~\bibnamefont{Hebenstreit}},
\bibnamefont{and} \bibinfo{author}{\bibfnamefont{O.~G.} \bibnamefont{Schmidt}},
\bibinfo{journal}{Advanced materials} \textbf{\bibinfo{volume}{28}},
\bibinfo{pages}{4084} (\bibinfo{year}{2016}).

\bibitem[{\citenamefont{Fournier-Bidoz   et~al.}(2005)\citenamefont{Fournier-Bidoz, Arsenault, Manners, and   Ozin}}]{fournier2005synthetic}
\bibinfo{author}{\bibfnamefont{S.}~\bibnamefont{Fournier-Bidoz}},
\bibinfo{author}{\bibfnamefont{A.~C.} \bibnamefont{Arsenault}},
\bibinfo{author}{\bibfnamefont{I.}~\bibnamefont{Manners}},
\bibnamefont{and} \bibinfo{author}{\bibfnamefont{G.~A.} \bibnamefont{Ozin}},
\bibinfo{journal}{Chemical Communications} pp. \bibinfo{pages}{441--443}
(\bibinfo{year}{2005}).

\bibitem[{\citenamefont{Marine et~al.}(2013)\citenamefont{Marine, Wheat, Ault,   and Posner}}]{marine2013diffusive}
\bibinfo{author}{\bibfnamefont{N.~A.} \bibnamefont{Marine}},
\bibinfo{author}{\bibfnamefont{P.~M.} \bibnamefont{Wheat}},
\bibinfo{author}{\bibfnamefont{J.}~\bibnamefont{Ault}},
\bibnamefont{and} \bibinfo{author}{\bibfnamefont{J.~D.} \bibnamefont{Posner}},
\bibinfo{journal}{Physical Review E} \textbf{\bibinfo{volume}{87}},
\bibinfo{pages}{052305} (\bibinfo{year}{2013}).

\bibitem[{\citenamefont{Takagi et~al.}(2013)\citenamefont{Takagi, Braunschweig,   Zhang, and Shelley}}]{takagi2013dispersion}
\bibinfo{author}{\bibfnamefont{D.}~\bibnamefont{Takagi}},
\bibinfo{author}{\bibfnamefont{A.~B.} \bibnamefont{Braunschweig}},
\bibinfo{author}{\bibfnamefont{J.}~\bibnamefont{Zhang}},
\bibnamefont{and} \bibinfo{author}{\bibfnamefont{M.~J.} \bibnamefont{Shelley}},
\bibinfo{journal}{Physical review letters} \textbf{\bibinfo{volume}{110}},
\bibinfo{pages}{038301} (\bibinfo{year}{2013}).

\bibitem[{\citenamefont{Buzhardt and   Tallapragada}(2021)}]{buzhardt2021magnetically}
\bibinfo{author}{\bibfnamefont{J.}~\bibnamefont{Buzhardt}}
\bibnamefont{and} \bibinfo{author}{\bibfnamefont{P.}~\bibnamefont{Tallapragada}},
\bibinfo{journal}{ASME Letters in Dynamic Systems and Control}
\textbf{\bibinfo{volume}{1}} (\bibinfo{year}{2021}).

\bibitem[{\citenamefont{Liebchen and L{"o}wen}(2019)}]{liebchen2019optimal}
\bibinfo{author}{\bibfnamefont{B.}~\bibnamefont{Liebchen}}
\bibnamefont{and} \bibinfo{author}{\bibfnamefont{H.}~\bibnamefont{L{"o}wen}},
\bibinfo{journal}{EPL (Europhysics Letters)} \textbf{\bibinfo{volume}{127}},
\bibinfo{pages}{34003} (\bibinfo{year}{2019}).

\bibitem[{\citenamefont{van Teeffelen and L{"o}wen}(2008)}]{van2008dynamics}
\bibinfo{author}{\bibfnamefont{S.}~\bibnamefont{van Teeffelen}}
\bibnamefont{and} \bibinfo{author}{\bibfnamefont{H.}~\bibnamefont{L{"o}wen}},
\bibinfo{journal}{Physical Review E} \textbf{\bibinfo{volume}{78}},
\bibinfo{pages}{020101} (\bibinfo{year}{2008}).

\bibitem[{\citenamefont{Muratore-Ginanneschi and   Schwieger}(2014)}]{muratore2014nanomechanical}
\bibinfo{author}{\bibfnamefont{P.}~\bibnamefont{Muratore-Ginanneschi}}
\bibnamefont{and} \bibinfo{author}{\bibfnamefont{K.}~\bibnamefont{Schwieger}},
\bibinfo{journal}{Physical Review E} \textbf{\bibinfo{volume}{90}},
\bibinfo{pages}{060102} (\bibinfo{year}{2014}).

\bibitem[{\citenamefont{Nemoto et~al.}(2019)\citenamefont{Nemoto, Fodor, Cates, Jack, and Tailleur}}]{nemoto2019optimizing}
\bibinfo{author}{\bibfnamefont{T.}~\bibnamefont{Nemoto}},
\bibinfo{author}{\bibfnamefont{E.}~\bibnamefont{Fodor}},
\bibinfo{author}{\bibfnamefont{M.~E.} \bibnamefont{Cates}},
\bibinfo{author}{\bibfnamefont{R.~L.} \bibnamefont{Jack}},
\bibnamefont{and} \bibinfo{author}{\bibfnamefont{J.}~\bibnamefont{Tailleur}},
\bibinfo{journal}{Physical Review E} \textbf{\bibinfo{volume}{99}},
\bibinfo{pages}{022605} (\bibinfo{year}{2019}).

\bibitem[{\citenamefont{Pietzonka et~al.}(2019)\citenamefont{Pietzonka, Fodor,   Lohrmann, Cates, and Seifert}}]{pietzonka2019autonomous}
\bibinfo{author}{\bibfnamefont{P.}~\bibnamefont{Pietzonka}},
\bibinfo{author}{\bibfnamefont{E.}~\bibnamefont{Fodor}},
\bibinfo{author}{\bibfnamefont{C.}~\bibnamefont{Lohrmann}},
\bibinfo{author}{\bibfnamefont{M.~E.} \bibnamefont{Cates}},
\bibnamefont{and} \bibinfo{author}{\bibfnamefont{U.}~\bibnamefont{Seifert}},
\bibinfo{journal}{Physical Review X} \textbf{\bibinfo{volume}{9}},
\bibinfo{pages}{041032} (\bibinfo{year}{2019}).

\bibitem[{\citenamefont{Seifert}(2012)}]{seifert2012stochastic} \bibinfo{author}{\bibfnamefont{U.}~\bibnamefont{Seifert}},
\bibinfo{journal}{Reports on progress in physics} \textbf{\bibinfo{volume}{75}},
\bibinfo{pages}{126001} (\bibinfo{year}{2012}).

\bibitem[{\citenamefont{Seifert}(2008)}]{seifert2008stochastic} \bibinfo{author}{\bibfnamefont{U.}~\bibnamefont{Seifert}},
\bibinfo{journal}{The European Physical Journal B} \textbf{\bibinfo{volume}{64}},
\bibinfo{pages}{423} (\bibinfo{year}{2008}).

\bibitem[{\citenamefont{Seifert}(2019)}]{seifert2019stochastic} \bibinfo{author}{\bibfnamefont{U.}~\bibnamefont{Seifert}},
\bibinfo{journal}{Annual Review of Condensed Matter Physics} \textbf{\bibinfo{volume}{10}},
\bibinfo{pages}{171} (\bibinfo{year}{2019}).

\bibitem[{\citenamefont{Jarzynski}(2011)}]{jarzynski2011equalities}
\bibinfo{author}{\bibfnamefont{C.}~\bibnamefont{Jarzynski}},
\bibinfo{journal}{Annu. Rev. Condens. Matter Phys.} \textbf{\bibinfo{volume}{2}},
\bibinfo{pages}{329} (\bibinfo{year}{2011}).

\bibitem[{\citenamefont{Barato and Seifert}(2015)}]{barato2015thermodynamic}
\bibinfo{author}{\bibfnamefont{A.~C.} \bibnamefont{Barato}}
\bibnamefont{and} \bibinfo{author}{\bibfnamefont{U.}~\bibnamefont{Seifert}},
\bibinfo{journal}{Physical review letters} \textbf{\bibinfo{volume}{114}},
\bibinfo{pages}{158101} (\bibinfo{year}{2015}).

\bibitem[{\citenamefont{Hwang and Hyeon}(2018)}]{hwang2018energetic}
\bibinfo{author}{\bibfnamefont{W.}~\bibnamefont{Hwang}}
\bibnamefont{and} \bibinfo{author}{\bibfnamefont{C.}~\bibnamefont{Hyeon}},
\bibinfo{journal}{The journal of physical chemistry letters} \textbf{\bibinfo{volume}{9}},
\bibinfo{pages}{513} (\bibinfo{year}{2018}).

\bibitem[{\citenamefont{Fodor et~al.}(2016)\citenamefont{Fodor, Nardini, Cates,
  Tailleur, Visco, and van Wijland}}]{fodor2016far}
\bibinfo{author}{\bibfnamefont{{\'E}.}~\bibnamefont{Fodor}},
  \bibinfo{author}{\bibfnamefont{C.}~\bibnamefont{Nardini}},
  \bibinfo{author}{\bibfnamefont{M.~E.} \bibnamefont{Cates}},
  \bibinfo{author}{\bibfnamefont{J.}~\bibnamefont{Tailleur}},
  \bibinfo{author}{\bibfnamefont{P.}~\bibnamefont{Visco}}, \bibnamefont{and}
  \bibinfo{author}{\bibfnamefont{F.}~\bibnamefont{van Wijland}},
  \bibinfo{journal}{Physical review letters} \textbf{\bibinfo{volume}{117}},
  \bibinfo{pages}{038103} (\bibinfo{year}{2016}).

\bibitem[{\citenamefont{Pearce and Giomi}(2016)}]{pearce2016linear}
\bibinfo{author}{\bibfnamefont{D.~J.} \bibnamefont{Pearce}} \bibnamefont{and}
  \bibinfo{author}{\bibfnamefont{L.}~\bibnamefont{Giomi}},
  \bibinfo{journal}{Physical Review E} \textbf{\bibinfo{volume}{94}},
  \bibinfo{pages}{022612} (\bibinfo{year}{2016}).

\bibitem[{\citenamefont{Kyriakopoulos et~al.}(2016)\citenamefont{Kyriakopoulos,
  Ginelli, and Toner}}]{kyriakopoulos2016leading}
\bibinfo{author}{\bibfnamefont{N.}~\bibnamefont{Kyriakopoulos}},
  \bibinfo{author}{\bibfnamefont{F.}~\bibnamefont{Ginelli}}, \bibnamefont{and}
  \bibinfo{author}{\bibfnamefont{J.}~\bibnamefont{Toner}},
  \bibinfo{journal}{New Journal of Physics} \textbf{\bibinfo{volume}{18}},
  \bibinfo{pages}{073039} (\bibinfo{year}{2016}).

\bibitem[{\citenamefont{Mandal et~al.}(2017)\citenamefont{Mandal, Klymko, and
  DeWeese}}]{mandal2017entropy}
\bibinfo{author}{\bibfnamefont{D.}~\bibnamefont{Mandal}},
  \bibinfo{author}{\bibfnamefont{K.}~\bibnamefont{Klymko}}, \bibnamefont{and}
  \bibinfo{author}{\bibfnamefont{M.~R.} \bibnamefont{DeWeese}},
  \bibinfo{journal}{Physical review letters} \textbf{\bibinfo{volume}{119}},
  \bibinfo{pages}{258001} (\bibinfo{year}{2017}).

\bibitem[{\citenamefont{Pietzonka and Seifert}(2017)}]{pietzonka2017entropy}
\bibinfo{author}{\bibfnamefont{P.}~\bibnamefont{Pietzonka}} \bibnamefont{and}
  \bibinfo{author}{\bibfnamefont{U.}~\bibnamefont{Seifert}},
  \bibinfo{journal}{Journal of Physics A: Mathematical and Theoretical}
  \textbf{\bibinfo{volume}{51}}, \bibinfo{pages}{01LT01}
  (\bibinfo{year}{2017}).

\bibitem[{\citenamefont{Shankar and Marchetti}(2018)}]{shankar2018hidden}
\bibinfo{author}{\bibfnamefont{S.}~\bibnamefont{Shankar}} \bibnamefont{and}
  \bibinfo{author}{\bibfnamefont{M.~C.} \bibnamefont{Marchetti}},
  \bibinfo{journal}{Physical Review E} \textbf{\bibinfo{volume}{98}},
  \bibinfo{pages}{020604} (\bibinfo{year}{2018}).

\bibitem[{\citenamefont{Pietzonka et~al.}(2019)\citenamefont{Pietzonka, Fodor,
  Lohrmann, Cates, and Seifert}}]{pietzonka2019autonomous}
\bibinfo{author}{\bibfnamefont{P.}~\bibnamefont{Pietzonka}},
  \bibinfo{author}{\bibfnamefont{{\'E}.}~\bibnamefont{Fodor}},
  \bibinfo{author}{\bibfnamefont{C.}~\bibnamefont{Lohrmann}},
  \bibinfo{author}{\bibfnamefont{M.~E.} \bibnamefont{Cates}}, \bibnamefont{and}
  \bibinfo{author}{\bibfnamefont{U.}~\bibnamefont{Seifert}},
  \bibinfo{journal}{Physical Review X} \textbf{\bibinfo{volume}{9}},
  \bibinfo{pages}{041032} (\bibinfo{year}{2019}).

\bibitem[{\citenamefont{Nemoto et~al.}(2019)\citenamefont{Nemoto, Fodor, Cates,
  Jack, and Tailleur}}]{nemoto2019optimizing}
\bibinfo{author}{\bibfnamefont{T.}~\bibnamefont{Nemoto}},
  \bibinfo{author}{\bibfnamefont{{\'E}.}~\bibnamefont{Fodor}},
  \bibinfo{author}{\bibfnamefont{M.~E.} \bibnamefont{Cates}},
  \bibinfo{author}{\bibfnamefont{R.~L.} \bibnamefont{Jack}}, \bibnamefont{and}
  \bibinfo{author}{\bibfnamefont{J.}~\bibnamefont{Tailleur}},
  \bibinfo{journal}{Physical Review E} \textbf{\bibinfo{volume}{99}},
  \bibinfo{pages}{022605} (\bibinfo{year}{2019}).

\bibitem[{\citenamefont{Dabelow et~al.}(2019)\citenamefont{Dabelow, Bo, and
  Eichhorn}}]{dabelow2019irreversibility}
\bibinfo{author}{\bibfnamefont{L.}~\bibnamefont{Dabelow}},
  \bibinfo{author}{\bibfnamefont{S.}~\bibnamefont{Bo}}, \bibnamefont{and}
  \bibinfo{author}{\bibfnamefont{R.}~\bibnamefont{Eichhorn}},
  \bibinfo{journal}{Physical Review X} \textbf{\bibinfo{volume}{9}},
  \bibinfo{pages}{021009} (\bibinfo{year}{2019}).

\bibitem[{\citenamefont{Burkholder and
  Brady}(2019)}]{burkholder2019fluctuation}
\bibinfo{author}{\bibfnamefont{E.W.} \bibnamefont{Burkholder}}
  \bibnamefont{and} \bibinfo{author}{\bibfnamefont{J.F.} \bibnamefont{Brady}},
  \bibinfo{journal}{The Journal of chemical physics}
  \textbf{\bibinfo{volume}{150}}, \bibinfo{pages}{184901}
  (\bibinfo{year}{2019}).

\bibitem[{\citenamefont{Bonilla}(2019)}]{bonilla2019active}
\bibinfo{author}{\bibfnamefont{L.L.} \bibnamefont{Bonilla}},
  \bibinfo{journal}{Physical Review E} \textbf{\bibinfo{volume}{100}},
  \bibinfo{pages}{022601} (\bibinfo{year}{2019}).

\bibitem[{\citenamefont{Szamel}(2019)}]{szamel2019stochastic} \bibinfo{author}{\bibfnamefont{G.}~\bibnamefont{Szamel}},
\bibinfo{journal}{Physical Review E} \textbf{\bibinfo{volume}{100}},
\bibinfo{pages}{050603} (\bibinfo{year}{2019}).

\bibitem[{\citenamefont{Cao et~al.}(2020)\citenamefont{Cao, Jiang, and   Hou}}]{cao2020design}
\bibinfo{author}{\bibfnamefont{Z.}~\bibnamefont{Cao}}, \bibinfo{author}{\bibfnamefont{H.}~\bibnamefont{Jiang}},
\bibnamefont{and} \bibinfo{author}{\bibfnamefont{Z.}~\bibnamefont{Hou}},
\bibinfo{journal}{Physical Review Research} \textbf{\bibinfo{volume}{2}},
\bibinfo{pages}{043331} (\bibinfo{year}{2020}).

\bibitem[{\citenamefont{Cao et~al.}(2015)\citenamefont{Cao, Wang, Ouyang, and   Tu}}]{cao2015free}
\bibinfo{author}{\bibfnamefont{Y.}~\bibnamefont{Cao}}, \bibinfo{author}{\bibfnamefont{H.}~\bibnamefont{Wang}},
\bibinfo{author}{\bibfnamefont{Q.}~\bibnamefont{Ouyang}},
\bibnamefont{and} \bibinfo{author}{\bibfnamefont{Y.}~\bibnamefont{Tu}},
\bibinfo{journal}{Nature Physics} \textbf{\bibinfo{volume}{11}},
\bibinfo{pages}{772} (\bibinfo{year}{2015}).

\bibitem[{\citenamefont{Horowitz and   Gingrich}(2019)}]{horowitz2019thermodynamic}
\bibinfo{author}{\bibfnamefont{J.~M.} \bibnamefont{Horowitz}}
\bibnamefont{and} \bibinfo{author}{\bibfnamefont{T.~R.} \bibnamefont{Gingrich}},
\bibinfo{journal}{Nature Physics} pp. \bibinfo{pages}{1--6}
(\bibinfo{year}{2019}).

\bibitem[{\citenamefont{Gingrich et~al.}(2016)\citenamefont{Gingrich, Horowitz,   Perunov, and England}}]{gingrich2016dissipation}
\bibinfo{author}{\bibfnamefont{T.~R.} \bibnamefont{Gingrich}},
\bibinfo{author}{\bibfnamefont{J.~M.} \bibnamefont{Horowitz}},
\bibinfo{author}{\bibfnamefont{N.}~\bibnamefont{Perunov}},
\bibnamefont{and} \bibinfo{author}{\bibfnamefont{J.~L.} \bibnamefont{England}},
\bibinfo{journal}{Physical review letters} \textbf{\bibinfo{volume}{116}},
\bibinfo{pages}{120601} (\bibinfo{year}{2016}).

\bibitem[{\citenamefont{Pietzonka et~al.}(2016)\citenamefont{Pietzonka, Barato,   and Seifert}}]{pietzonka2016universal}
\bibinfo{author}{\bibfnamefont{P.}~\bibnamefont{Pietzonka}},
\bibinfo{author}{\bibfnamefont{A.~C.} \bibnamefont{Barato}},
\bibnamefont{and} \bibinfo{author}{\bibfnamefont{U.}~\bibnamefont{Seifert}},
\bibinfo{journal}{Physical Review E} \textbf{\bibinfo{volume}{93}},
\bibinfo{pages}{052145} (\bibinfo{year}{2016}).

\bibitem[{\citenamefont{Pigolotti et~al.}(2017)\citenamefont{Pigolotti, Neri,   Rold{\'a}n, and J{"u}licher}}]{pigolotti2017generic}
\bibinfo{author}{\bibfnamefont{S.}~\bibnamefont{Pigolotti}},
\bibinfo{author}{\bibfnamefont{I.}~\bibnamefont{Neri}},
\bibinfo{author}{\bibfnamefont{E.}~\bibnamefont{Rold{\'a}n}},
\bibnamefont{and} \bibinfo{author}{\bibfnamefont{F.}~\bibnamefont{J{"u}licher}},
\bibinfo{journal}{Physical review letters} \textbf{\bibinfo{volume}{119}},
\bibinfo{pages}{140604} (\bibinfo{year}{2017}).

\bibitem[{\citenamefont{Proesmans and Van~den   Broeck}(2017)}]{proesmans2017discrete}
\bibinfo{author}{\bibfnamefont{K.}~\bibnamefont{Proesmans}}
\bibnamefont{and} \bibinfo{author}{\bibfnamefont{C.}~\bibnamefont{Van~den
Broeck}}, \bibinfo{journal}{EPL (Europhysics Letters)} \textbf{\bibinfo{volume}{119}},
\bibinfo{pages}{20001} (\bibinfo{year}{2017}).

\bibitem[{\citenamefont{Dechant and   Sasa}(2018{\natexlab{a}})}]{dechant2018entropic}
\bibinfo{author}{\bibfnamefont{A.}~\bibnamefont{Dechant}}
\bibnamefont{and} \bibinfo{author}{\bibfnamefont{S.-i.} \bibnamefont{Sasa}},
\bibinfo{journal}{Physical Review E} \textbf{\bibinfo{volume}{97}},
\bibinfo{pages}{062101} (\bibinfo{year}{2018}{\natexlab{a}}).

\bibitem[{\citenamefont{Van~Vu and Hasegawa}(2019)}]{van2019uncertainty}
\bibinfo{author}{\bibfnamefont{T.}~\bibnamefont{Van~Vu}}
\bibnamefont{and} \bibinfo{author}{\bibfnamefont{Y.}~\bibnamefont{Hasegawa}},
\bibinfo{journal}{Physical Review E} \textbf{\bibinfo{volume}{100}},
\bibinfo{pages}{032130} (\bibinfo{year}{2019}).

\bibitem[{\citenamefont{Hasegawa and Van~Vu}(2019)}]{hasegawa2019fluctuation}
\bibinfo{author}{\bibfnamefont{Y.}~\bibnamefont{Hasegawa}}
\bibnamefont{and} \bibinfo{author}{\bibfnamefont{T.}~\bibnamefont{Van~Vu}},
\bibinfo{journal}{Physical Review Letters} \textbf{\bibinfo{volume}{123}},
\bibinfo{pages}{110602} (\bibinfo{year}{2019}).

\bibitem[{\citenamefont{Dechant and   Sasa}(2018{\natexlab{b}})}]{dechant2018current}
\bibinfo{author}{\bibfnamefont{A.}~\bibnamefont{Dechant}}
\bibnamefont{and} \bibinfo{author}{\bibfnamefont{S.-i.} \bibnamefont{Sasa}},
\bibinfo{journal}{Journal of Statistical Mechanics: Theory and Experiment}
\textbf{\bibinfo{volume}{2018}}, \bibinfo{pages}{063209} (\bibinfo{year}{2018}{\natexlab{b}}).

\bibitem[{\citenamefont{Farage et~al.}(2015)\citenamefont{Farage, Krinninger,
  and Brader}}]{farage2015effective}
\bibinfo{author}{\bibfnamefont{T.~F.} \bibnamefont{Farage}},
  \bibinfo{author}{\bibfnamefont{P.}~\bibnamefont{Krinninger}},
  \bibnamefont{and} \bibinfo{author}{\bibfnamefont{J.~M.}
  \bibnamefont{Brader}}, \bibinfo{journal}{Physical Review E}
  \textbf{\bibinfo{volume}{91}}, \bibinfo{pages}{042310}
  (\bibinfo{year}{2015}).

\bibitem[{\citenamefont{Zhang et~al.}(2020)\citenamefont{Zhang, Cao, Ouyang,
  and Tu}}]{2020The}
\bibinfo{author}{\bibfnamefont{D.}~\bibnamefont{Zhang}},
  \bibinfo{author}{\bibfnamefont{Y.}~\bibnamefont{Cao}},
  \bibinfo{author}{\bibfnamefont{Q.}~\bibnamefont{Ouyang}}, \bibnamefont{and}
  \bibinfo{author}{\bibfnamefont{Y.}~\bibnamefont{Tu}},
  \bibinfo{journal}{Nature Physics}  (\bibinfo{year}{2020}).

\bibitem[{\citenamefont{Lee et~al.}(2018)\citenamefont{Lee, Hyeon, and
  Jo}}]{lee2018thermodynamic}
\bibinfo{author}{\bibfnamefont{S.}~\bibnamefont{Lee}},
  \bibinfo{author}{\bibfnamefont{C.}~\bibnamefont{Hyeon}}, \bibnamefont{and}
  \bibinfo{author}{\bibfnamefont{J.}~\bibnamefont{Jo}},
  \bibinfo{journal}{Physical Review E} \textbf{\bibinfo{volume}{98}},
  \bibinfo{pages}{032119} (\bibinfo{year}{2018}).

\bibitem[{\citenamefont{Sokolov}(2012)}]{sokolov2012models}
\bibinfo{author}{\bibfnamefont{I.~M.} \bibnamefont{Sokolov}},
  \bibinfo{journal}{Soft Matter} \textbf{\bibinfo{volume}{8}},
  \bibinfo{pages}{9043} (\bibinfo{year}{2012}).

\end{thebibliography}

\expandafter\ifx\csname natexlab\endcsname\relax
\global\long\def\natexlab#1{#1}%
\fi \expandafter\ifx\csname bibnamefont\endcsname\relax
\global\long\def\bibnamefont#1{#1}%
\fi \expandafter\ifx\csname bibfnamefont\endcsname\relax
\global\long\def\bibfnamefont#1{#1}%
\fi \expandafter\ifx\csname citenamefont\endcsname\relax
\global\long\def\citenamefont#1{#1}%
\fi \expandafter\ifx\csname url\endcsname\relax
\global\long\def\url#1{\texttt{#1}}%
\fi \expandafter\ifx\csname urlprefix\endcsname\relax
\global\long\def\urlprefix{URL }%
\fi \providecommand{\bibinfo}[2]{#2} \providecommand{\eprint}[2][]{\url{#2}}

\end{document}